\pdfpageattr{/Group << /S /Transparency /I true /CS /DeviceRGB>>}
\documentclass[aps,prc,reprint,superscriptaddress,showpacs,amsmath,nofootinbib]{revtex4-1}
\usepackage[T1]{fontenc}
\usepackage[utf8]{inputenc}

\usepackage{isotope}
\usepackage{xcolor}
\usepackage{graphicx}
\usepackage[load-configurations=abbreviations]{siunitx}
\usepackage{braket}
\usepackage{mathtools}
\usepackage{isotope}
\usepackage{ifpdf}

\usepackage{txfonts}
\usepackage{booktabs}

\let\rtxappendix\appendix
\usepackage{hyperref}
\usepackage{cleveref}
\let\Hyappendix\appendix
\AtBeginDocument{\renewcommand*{\appendix}{\Hyappendix\rtxappendix}}

\ifpdf\else\renewcommand{\href}[2]{#2}\fi

\crefname{equation}{}{}
\Crefname{equation}{Equation}{Equations}
\crefname{figure}{Fig.}{Figs.}
\crefname{table}{Table}{Tables}

\makeatletter
\DeclareUnicodeCharacter{014C}{\@tabacckludge=O} 
\DeclareUnicodeCharacter{014D}{\@tabacckludge=o} 
\makeatother

\DeclareSIUnit{\GeV}{\giga\electronvolt}
\DeclareSIUnit{\MeV}{\mega\electronvolt}
\DeclareSIUnit{\keV}{\kilo\electronvolt}
\DeclareSIUnit{\fm}{\femto\meter}
\DeclareSIUnit{\Msol}{\ensuremath{M_\odot}}
\DeclareSIUnit{\MeVc}{\MeV\per\text{\ensuremath{c}}}

\newcommand{\comm}[2]{\lbrack #1,#2\rbrack}
\newcommand{\kronecker}[2]{\delta^{#1}_{#2}}
\newcommand{\clebsch}[6]{\ensuremath{C_{#1#2,#3#4}^{#5#6}}} 
\newcommand{\sixj}[6]{\begingroup\setlength{\arraycolsep}{0.2em}\begin{Bmatrix} #1 & #2 & #3 \\ #4 & #5 & #6 \end{Bmatrix}\endgroup}
\newcommand{\ninej}[9]{\begingroup\setlength{\arraycolsep}{0.2em}\begin{Bmatrix} #1 & #2 & #3 \\ #4 & #5 & #6 \\ #7 & #8 & #9 \end{Bmatrix}\endgroup}
\newcommand{\hob}[6]{\langle\!\langle #1,#2\mspace{4mu}|\mspace{4mu}#3,#4:#5\rangle\!\rangle_{#6}}

\newcommand{\Tcoeff}[9]{T^{[(\JTred #1\JTred #2)#4#5,\JTred #3]#6}_{(#7_\text{cm}#8_\text{cm},\JTred #9)#6}}
\newcommand{\Tncoeff}[9]{T'^{[(\JTred #1\JTred #2)#4,\JTred #3]#6}_{(#7_\text{cm}#8_\text{cm},\JTred #9)#6}} 

\newcommand{\keta}[1]{\ket{#1}_a}

\newcommand{\pcl}{\chi}

\newcommand{\Strng}{\mathcal{S}}
\newcommand{\mt}{\tau}
\newcommand{\jhat}{\hat{\jmath}}
\newcommand{\Nmax}{\ensuremath{N_\text{max}}}

\newcommand*{\op}[1]{\boldsymbol{#1}}
\newcommand{\Ham}{\op{H}}
\newcommand{\Tint}{\op{T}_{\mspace{-5mu}\text{int}}}

\newcommand*{\vect}[1]{\vec{#1}}

\newcommand{\mindent}{\mspace{18mu}}

\newcommand{\Jcal}{\mathcal{J}}
\newcommand{\Mcal}{\mathcal{M}}
\newcommand{\Xcal}{\mathcal{X}}

\newcommand{\JTred}[1]{\tilde{#1}}

\newcommand{\EMthreeL}{N\textsuperscript3LO\textsubscript{EM}+N\textsuperscript2LO\textsubscript{L}}
\newcommand{\EMNfourNL}{N\textsuperscript4LO\textsubscript{EMN}+N\textsuperscript2LO\textsubscript{NL}}
\newcommand{\EMthreeNL}{N\textsuperscript3LO\textsubscript{EM}+N\textsuperscript2LO\textsubscript{NL}}

\newcommand{\coloronline}{(color online) }

\allowdisplaybreaks[1]

\begin{document}

\title{Similarity renormalization group evolution of hypernuclear Hamiltonians}

\author{Roland Wirth}
\email{roland.wirth@physik.tu-darmstadt.de}
\affiliation{Institut für Kernphysik -- Theoriezentrum, TU Darmstadt, Schlossgartenstr. 2, 64289 Darmstadt, Germany}
\affiliation{Facility for Rare Isotope Beams, Michigan State University, East Lansing, Michigan 48824, USA}

\author{Robert Roth}
\email{robert.roth@physik.tu-darmstadt.de}
\affiliation{Institut für Kernphysik -- Theoriezentrum, TU Darmstadt, Schlossgartenstr. 2, 64289 Darmstadt, Germany}

\date{\today}

\begin{abstract}
Unitary transformations of a Hamiltonian generally induce interaction terms beyond the particle rank present in the untransformed Hamiltonian that have to be captured and included in a many-body calculation.
In systems with strangeness such as hypernuclei, the three-body terms induced by the hyperon-nucleon interaction are strong, so their inclusion is crucial.

We present in detail a procedure for computing hyperon-nucleon-nucleon interaction terms that are induced during a similarity renormalization group (SRG) flow.
The SRG is carried out in a basis spanned by antisymmetric harmonic-oscillator states with respect to three-body Jacobi coordinates.
We discuss basis construction, antisymmetrization, numerical evaluation of the flow equations, and separation of the genuine three-body terms.

We then use the hypernuclear no-core shell model with SRG-evolved Hamiltonians, addressing the sensitivity of hypernuclear states and hyperon separation energies to changes in the nucleonic Hamiltonian by example of \isotope[7][\Lambda]{Li}, \isotope[9][\Lambda]{Be}, \isotope[11][\Lambda]{B}, \isotope[13][\Lambda]{C}, and the hyper-helium chain.
We also present a survey of the hyper-hydrogen chain, exploring the structure of hypernuclei with extreme neutron-proton asymmetries.
\end{abstract}

\maketitle

\section{Introduction}

The understanding of strangeness in finite and infinite strongly-interacting systems is key to not only gaining insight into the low-energy limit of the strong interaction itself, but also to understanding the structure of neutron stars \cite{Hiyama2009,Chatterjee2016,Gal2016,Vidana2018}.
Recently, we developed the hypernuclear no-core shell model (NCSM) \cite{Wirth2014,Wirth2018a}, the first \emph{ab initio} method able to calculate hypernuclei beyond the $s$ shell with nonlocal interactions, such as those derived from chiral effective field theory \cite{Polinder2006,Haidenbauer2013}.
With that, we have established a link between the low-energy effective field theory of QCD and the phenomenology of hypernuclei.

The NCSM and all other basis-expansion approaches to the nuclear many-body problem rely on model spaces spanned by finite basis sets. The convergence of many-body observables with increasing model-space size contributes to the theoretical uncertainties and eventually limits the range of ab initio calculations in terms of mass number. Therefore, the acceleration of convergence via unitary or similarity transformations of the Hamiltonian and other relevant operators is a key ingredient in ab initio nuclear and hypernuclear structure theory.
The similarity renormalization group (SRG) \cite{Gazek1993,Wegner1994,*Wegner2000} has proven to be a very versatile and effective tool to achieve this convergence acceleration.
The tradeoff that comes with using unitary transformations is that these transformations induce many-body interactions beyond those that are present in the initial Hamiltonian.
Contrary to other methods, the SRG allows for the explicit computation of induced many-body terms and of consistently transformed operators in a conceptually straight-forward manner.

In our previous works we have presented the first NCSM calculations for $p$\nobreakdash-shell hypernuclei with chiral two- and three-baryon interactions \cite{Wirth2014,Wirth2018} and we have demonstrated that hyperon-nucleon-nucleon (YNN) terms induced by the SRG transformation are strong and cannot be neglected when working with transformed interactions \cite{Wirth2016}. The size of SRG-induced YNN interactions is remarkable and highlights a special feature of the hyperon-nucleon (YN) interactions, the conversion of a $\Lambda$ to a $\Sigma$ hyperon in an interaction process with a nucleon.
We have shown that the elimination of this conversion, i.e. the decoupling of $\Lambda$ and $\Sigma$ channels, through an SRG evolution leads to strong repulsive $\Lambda$NN interactions \cite{Wirth2016}. As a consequence, models for hyperons in matter that only include the $\Lambda$ hyperon and omit the $\Lambda$-$\Sigma$ conversion have to include strong repulsive $\Lambda$NN forces. This has direct impact on the hyperon puzzle in neutron-star physics \cite{Wirth2016,Chatterjee2016,Lonardoni2015}.

In Refs.\ \cite{Wirth2016,Wirth2018} we have focused on the applications and the physics discussion of the hypernuclear SRG. In the present paper we provide a detailed discussion of the formalism and, particularly, the extension of the SRG to the YNN three-baryon sector with all elements necessary for the practical implementation. In addition we present new calculations for light hypernuclei, including the hydrogen and helium isotopic chains up to the driplines, and investigate the impact of the nucleonic part of the Hamiltonian on hypernuclear spectra.

This paper is organized as follows:
\Cref{sec:ynn} considers the necessary steps for computing the YNN terms induced during the SRG evolution of a two-body Hamiltonian.
We give a short overview over the core ideas of the hypernuclear no-core shell model in \cref{sec:ncsm}.
In \cref{sec:results} we present NCSM calculations for a sample set of hypernuclei using state-of-the-art nucleonic Hamiltonians, which provide better saturation properties than the one used before.

\section{\label{sec:ynn}Similarity renormalization group}
Most models of baryon-baryon interactions display strong repulsion at short distances where the baryons overlap.
This repulsive core strongly couples two-baryon states with low and high relative momentum.
In order to accommodate these couplings, which is necessary to achieve convergence of the many-body calculation, large many-body model spaces are needed.
For all but the lightest systems, the sizes needed are beyond the capabilities of current high-performance computers.
To accelerate convergence and reduce the required model-space sizes, methods based on unitary transformations of the Hamiltonian have been devised that suppress the coupling between low- and high-momentum states.
A simple and very versatile method is the SRG.

\subsection{Formalism}
The SRG is a continuous unitary transformation $\op{H}(\alpha) = \op{U}^\dag(\alpha)\op{H}(0)\op{U}(\alpha)$ of a Hamiltonian, depending on a flow parameter $\alpha$.
It is governed by the flow equation
\begin{equation}
  \partial_\alpha \op{H}(\alpha) = \comm{\op{\eta}(\alpha)}{\op{H}(\alpha)}, \label{eq:ynn:flow equation}
\end{equation}
where $\partial_\alpha$ denotes the derivative with respect to $\alpha$.
The anti-Hermitian generator $\op{\eta}(\alpha) = (\partial_\alpha \op{U}^\dag(\alpha))\op{U}(\alpha)$ can be chosen freely to achieve a desired behavior.
A very general and convenient ansatz for $\op{\eta}(\alpha)$ is a commutator $\comm{\op{\Gamma}}{\op{H}(\alpha)}$ so that the flow stops when the Hamiltonian commutes with the Hermitian operator $\op{\Gamma}$.
In nuclear physics, we commonly use $\op{\Gamma}=m_N^2 \Tint$, the intrinsic kinetic energy, which drives the Hamiltonian to band-diagonal form in HO basis.
The squared nucleon mass $m^2_N$ sets the unit of $\alpha$ to \si{\fm\tothe4}.

Other observables can be evolved consistently by solving the same differential equation \cref{eq:ynn:flow equation} for the observable.
Note, however, that the generator depends on the Hamiltonian, so that the observables have to be evolved simultaneously.
When considering multiple observables, it is more economical to compute the unitary transformation $\op{U}$ by solving
\begin{equation}
  \partial_\alpha \op{U}(\alpha) = - \op{U}(\alpha)\op{\eta}(\alpha)
\end{equation}
while evolving the Hamiltonian and transform the observables afterwards.

\subsection{Evolution in two-body space}
The abstract operator differential equation \cref{eq:ynn:flow equation} needs to be converted to a flow equation for matrix elements that can be evaluated numerically.
We begin by discussing the evolution in two-body space to show the necessary steps without introducing the complexity of a three-body system.

For a general $A$-body system with charge $Q$ and strange\-ness\footnote{Strangeness is defined as the number of anti-strange minus the number of strange quarks so hypernuclei have $\Strng < 0$.} $\Strng$, the starting point of the evolution is a Hamiltonian $\op{H}=\Tint + \Delta\op{M} + \op{V}_\text{NN} + \op{V}_\text{YN} + \op{V}_\text{NNN} + \dotsb$.
The first term is the intrinsic kinetic energy
\begin{equation}
  \Tint = \sum_{i=1}^A \frac{\vect{\op{p}}^2_i}{2\op{m}_i} - \op{T}_{c.m.},
\end{equation}
with single-particle momenta $\vect{\op{p}}_i$, masses $\op{m}_i$, and center-of-mass kinetic energy $\op{T}_{c.m.}$.
The second term,
\begin{equation}
  \Delta\op{M} = \sum_{i=1}^A \op{m}_i - M_0,
\end{equation}
where $M_0 = (A - Q - \lvert\Strng\rvert) m_n + Q m_p + \lvert\Strng\rvert m_\Lambda$ is the total rest mass of the noninteracting system of protons, neutrons, and $\Lambda$ hyperons, accounts for the higher rest mass of the $\Sigma$ hyperon.
The remaining terms are two-body NN and YN interactions, the three-nucleon (NNN) interaction, and higher interaction terms that are neglected.
For the evolution in two-body space, the NNN interaction can also be omitted.

To carry out the evolution, we take matrix elements of the Hamiltonian between harmonic-oscillator (HO) states.
These states are defined with respect to the center-of-mass and relative Jacobi coordinates $\vect{\xi}_0$ and $\vect{\xi}_1$,
\begin{align}
  \vect{\xi}_0 &= \frac1{\sqrt{M_2}}(\sqrt{m_b} \vect{x}_a + \sqrt{m_2} \vect{x}_2) \label{eq:ynn:2b jacobi cm}\\
  \vect{\xi}_1 &= \frac1{\sqrt{M_2}}(\sqrt{m_b} \vect{x}_a - \sqrt{m_a} \vect{x}_b), \label{eq:ynn:2b jacobi rel}
\shortintertext{where}
  \vect{x}_i &= \sqrt{\frac{m_i}{m_N}} \vect{r}_i \\
  M_n &= \sum_{k=1}^n m_k.
\end{align}
The $m_i$ and $\vect{r}_i$ are the masses and positions of the particles, respectively, and the nucleon mass $m_N$ is used as a global scaling factor to make the $\vect{x}_i$ have units of length.
The Jacobi coordinates coincide with the ones commonly used for nucleonic systems if $m_i=m_N$ for all particles.

Since the interactions are translationally invariant, we can separate the center-of-mass degrees of freedom and define a basis of antisymmetric relative HO states $\keta{[nl,(s_as_b)S_{ab}]JM,\pcl_a\pcl_b}$, with $s_{a/b}$ the spins of the particles, coupled to total spin $S_{ab}$, the radial quantum number $n$ and orbital angular momentum $l$ of the relative motion, total angular momentum $J$ with projection $M$, and with $\pcl_{a/b} =\{\Strng t \mt\}$ denoting the species, i.e.\ strangeness, isospin and isospin projection, of the two particles.
To make this two-body basis finite, we introduce an energy truncation $e = 2n+l \le E_\text{2,max}$.
In the finite space spanned by the basis states, the SRG flow equation becomes an ordinary matrix differential equation that can be solved numerically.
Due to symmetries of the Hamiltonian, only certain subsets of states can produce non\-vanishing matrix elements, such that the Hamiltonian decomposes into blocks of equal total angular momentum $J$, charge $Q$ and strangeness $\Strng$, which can be evolved separately.
Also, the matrix elements are independent of the angular momentum projection $M$.
The SRG transformation on the two-body level is given in more detail in Ref.~\cite{Wirth2018a}.

An important aspect of the SRG evolution of the YN interaction in two-body space is the suppression of $\Lambda$-$\Sigma$ conversion matrix elements, i.e., for sufficiently large flow parameters, we can remove the $\Sigma$ degrees of freedom from the model space without affecting the low-lying states.
However, as shown in Ref.~\cite{Wirth2016}, the evolved two-body interaction is too attractive and massively overbinds hypernuclei compared to the unevolved one.
Since the evolution changes the eigenenergies of the many-body system, unitarity in the $A$-body system is violated and we have to include at least (repulsive) induced three-body terms in order to restore it.

\subsection{Evolution in three-body space}
The evolution in two-body space is, by construction, unable to capture induced terms beyond the two-body level.
To determine the induced three-body terms, the evolution has to be carried out in three-body space.
There, the initial Hamiltonian can already contain three-body forces.
However, the NNN and YNN forces act on the $\Strng=0$ and $\Strng=-1$ sectors of the three-particle Hilbert space, respectively, so only one of them needs to be included in the evolution.
Moreover, initial YNN forces do not enter the chiral expansion until N\textsuperscript2LO and are, thus, absent from the calculation presented here.

\subsubsection{Coordinate systems, basis sets, and transformations}
Contrary to the two-body system \cref{eq:ynn:2b jacobi cm,eq:ynn:2b jacobi rel} the three-body case has multiple sets of Jacobi coordinates, one of which is a straight-forward extension of the two-body ones:
\begin{align}
  \vect{\xi}_0 &= \frac1{\sqrt{M_3}}(\sqrt{m_a} \vect{x}_a + \sqrt{m_b} \vect{x}_b + \sqrt{m_c} \vect{x}_c)\\
  \vect{\xi}_1 &= \frac1{\sqrt{M_2}}(\sqrt{m_b} \vect{x}_a - \sqrt{m_a} \vect{x}_b) \\
  \vect{\xi}_2 &= \frac1{\sqrt{M_3}} \bigl(\sqrt{m_3}\vect{X}_{ab} - \sqrt{M_2}\vect{x}_c\bigr) \\
  \vect{X}_{ab} &= \frac1{\sqrt{M_2}} \bigl(\sqrt{m_a}\vect{x}_a + \sqrt{m_b}\vect{x}_b\bigr),
\end{align}
with the center-of-mass coordinate $\vect{X}_{ab}$ of particles $a$ and $b$ (which coincides with \cref{eq:ynn:2b jacobi cm}).
One can also define a different set of coordinates
\begin{align}
  \vect{\xi}'_0 &= \vect{\xi}_0 \\
  \vect{\xi}'_1 &= \frac{1}{\sqrt{m_a+m_c}}\bigl(\sqrt{m_c}\vect{x}_a - \sqrt{m_a}\vect{x}_c\bigr) \\
  \vect{\xi}'_2 &= \frac{1}{\sqrt{M_3}} \Biggl(\sqrt{\frac{m_b}{m_a+m_c}} \bigl(\sqrt{m_a}\vect{x}_a + \sqrt{m_c}\vect{x}_c\bigr) - \sqrt{m_a+m_c}\vect{x}_b\Biggr),
\end{align}
where the first coordinate is defined by the first and third particle.
This set is needed for antisymmetrization.

The $\vect{\xi}_i$, $\vect{\xi}'_i$, and $\vect{x}_i$ are connected via orthogonal transformations of the type
\begin{equation}\label{eq:ynn:orthogonal transformation}
  \begin{pmatrix}
    \vect{V} \\ \vect{v}
  \end{pmatrix}
  =
  \begin{pmatrix}
    \sqrt{\frac{d}{1+d}} & \sqrt{\frac{1}{1+d}} \\
    \sqrt{\frac{1}{1+d}} & -\sqrt{\frac{d}{1+d}}
  \end{pmatrix}
  \begin{pmatrix}
    \vect{v}_1 \\ \vect{v}_2
  \end{pmatrix},
\end{equation}
parametrized by the nonnegative number $d$, which relate two general pairs of coordinates $\vect{V},\vect{v}$ and $\vect{v}_1,\vect{v}_2$.
The transformation from $\vect{\xi}_i$ to the single-particle coordinates $\vect{x}_i$ is effected in two steps by first transforming $\vect{\xi}_0,\vect{\xi}_2$ to $\vect{X}_{ab},\vect{x}_c$ ($d_1=(m_a+m_b)/m_c$) and then $\vect{X}_{ab},\vect{\xi}_1$ to $\vect{x}_a,\vect{x}_b$ with $d_2=m_a/m_b$.
Transforming between the $\vect{\xi}'_i$ and $\vect{\xi}_i$ is needed to antisymmetrize the three-body states.
Since the center-of-mass coordinate is the same in both sets, it suffices to find the orthogonal transformation connecting $\vect{\xi}'_1,\vect{\xi}'_2$ and $\vect{\xi}_1,\vect{\xi}_2$, which after some algebra turns out to have $D=m_bm_c/(m_a(m_a+m_b+m_c))$.

For HO states, the overlap between states $\ket{(NL,nl)\lambda M_\lambda}$ and $\ket{(n_1l_1,n_2l_2)\lambda M_\lambda}$ defined with respect to two coordinate pairs $\vect{R},\vect{r}$ and $\vect{r}_1,\vect{r}_2$ related by an orthogonal transformation \cref{eq:ynn:orthogonal transformation} is given by a harmonic-oscillator bracket (HOB) $\hob{NL}{nl}{n_1l_1}{n_2l_2}{\lambda}{d}$.
The HOBs can be computed analytically, e.g., using the expressions given by \citet{Kamuntavicius2001}.

In the following we will work in a basis spanned by HO states with respect to the Jacobi coordinates $\vect{\xi}_i$, coupled to total angular momentum and isospin.
We choose to work in an isospin-coupled basis and neglect isospin breaking of the induced YNN terms because it is computationally too demanding to take this effect into account.

For the upcoming derivations we define a basis set that is antisymmetric only under exchange of the first two particles, indicated by a subscript ``$ab$'',
\begin{align}
    \MoveEqLeft\ket{(n_\text{cm}l_\text{cm},\alpha)JM}_{ab} = \notag\\
      &\bigl(2+2\kronecker{s_at_a\Strng_a}{s_bt_b\Strng_b}\bigr)^{-1/2} \bigl(\ket{(n_\text{cm}l_\text{cm},\alpha)JM} \notag\\*
      &\mindent - (-1)^{l_1+s_a+s_b-S_{ab}+t_a+t_b-T_{ab}}\ket{(n_\text{cm}l_\text{cm},\alpha[a\leftrightarrow b])JM}\bigr). \label{eq:ynn:alpha basis}
\end{align}
The intrinsic quantum numbers are collected into $\alpha = \{[n_1l_1,(s_as_b)S_{ab}]j_1,[n_2l_2,s_c]j_2\}\Jcal,[(\Strng_at_a\Strng_bt_b)T_{ab},\Strng_ct_c]TM_T$ and the notation $\alpha[a\leftrightarrow b]$ denotes the set where the quantum numbers of particles $a$ and $b$ have been exchanged.
The Kronecker delta $\kronecker{s_at_a\Strng_a}{s_bt_b\Strng_b}$ is short notation for $\kronecker{s_a}{s_b}\kronecker{t_a}{t_b}\kronecker{\Strng_a}{\Strng_b}$.

\subsubsection{Antisymmetrization}
Contrary to the two-body sector, antisymmetrizing a three-body state is not trivial except for product states.
We achieve antisymmetrization in the Jacobi HO basis through explicit projection onto the antisymmetric subspace.
The antisymmetric subspace is the space spanned by the eigenvectors of the antisymmetrizer $\mathcal{A}$ to eigenvalue $1$.
Diagonalizing the antisymmetrizer, represented as a matrix with respect to the partially-antisymmetric Jacobi HO basis \cref{eq:ynn:alpha basis}, gives a basis of the antisymmetric subspace in terms of these basis states.

The antisymmetrizer for a three-body system is
\begin{equation}
  \op{\mathcal{A}} = \tfrac1{3!} (\op{1} - \op{P}_{ab} - \op{P}_{ac} - \op{P}_{bc} + \op{P}_{bc}\op{P}_{ab} + \op{P}_{ac}\op{P}_{ab})
\end{equation}
and with respect to the $ab$-antisymmetric basis, exploiting that $\op{P}_{ac}=\op{P}_{ab}\op{P}_{bc}\op{P}_{ab}$ and the eigenvalue relation of $\op{P}_{ab}$, its matrix elements
\begin{align}
  \MoveEqLeft\prescript{}{ab}{\braket{(n_\text{cm}l_\text{cm},\alpha)JM|\op{\mathcal{A}}|(n'_\text{cm}l'_\text{cm},\alpha')J'M'}}_{ab} = \notag\\
    &\tfrac13 \prescript{}{ab}{\braket{(n_\text{cm}l_\text{cm},\alpha)JM|(\op{1} - 2\op{P}_{bc})|(n'_\text{cm}l'_\text{cm},\alpha')J'M'}}_{ab}
\end{align}
are trivially related to the matrix elements of of the transposition operator $\op{P}_{bc}$.
We separate the spin, isospin and spatial parts of the matrix element and consider them separately (for simplicity, we consider a nonantisymmetric matrix element and apply \cref{eq:ynn:alpha basis} to get the final result):
\begin{widetext}
\begin{align}
  \MoveEqLeft \braket{(n_\text{cm}l_\text{cm},\alpha)JM|\op{P}_{bc}|(n'_\text{cm}l'_\text{cm},\alpha')J'M'} =
\notag\\*&
  \sum_{\substack{m_\text{cm} \Mcal \\ m'_\text{cm} \Mcal'}}\sum_{\substack{L S \\ L' S'}}\sum_{\substack{M_L M_S \\ M'_L M'_S}}
    \jhat_1\jhat'_1\jhat_2\jhat'_2\hat{L}\hat{L}'\hat{S}\hat{S}'
    \ninej{l_1}{S_{ab}}{j_1}{l_2}{s_c}{j_2}{L}{S}{\Jcal}
    \ninej{l'_1}{S'_{ab}}{j'_1}{l'_2}{s'_c}{j'_2}{L'}{S'}{\Jcal'}
    \clebsch{l_\text{cm}}{m_\text{cm}}{\Jcal}{\Mcal}{J}{M}
    \clebsch{l'_\text{cm}}{m'_\text{cm}}{\Jcal'}{\Mcal'}{J'}{M'}
    \clebsch{L}{M_L}{S}{M_S}{\Jcal}{\Mcal}
    \clebsch{L'}{M'_L}{S'}{M'_S}{\Jcal'}{\Mcal'}
\notag\\&\hphantom{{}={}}\times
    \braket{[(s_as_b)S_{ab},s_c]SM_S|\op{P}_{bc}|[(s'_as'_b)S'_{ab},s'_c]S'M'_S}
    \braket{[(\Strng_at_a\Strng_bt_b)T_{ab},\Strng_ct_c]TM_T|\op{P}_{bc}|[(\Strng'_at'_a\Strng'_bt'_b)T'_{ab},\Strng'_ct'_c]T'M'_T}
\notag\\*&\hphantom{{}={}}\times
    \braket{n_\text{cm}l_\text{cm}m_\text{cm},(n_1l_1n_2l_2)LM_L|\op{P}_{bc}|n'_\text{cm}l'_\text{cm}m'_\text{cm},(n'_1l'_1n'_2l'_2)L'M'_L}.
  \label{eq:ynn:pbc matrix element 1}
\end{align}

The spin and isospin parts have the same structure apart from an additional constraint on the strangeness quantum numbers.
The application of the permutation changes the coupling order to a scheme where particles 1 and 3 are coupled first and particle 2 couples to the resulting spin.
The matrix element is
\begin{align}
  \braket{[(s_as_b)S_{ab},s_c]SM_S|\op{P}_{bc}|[(s'_as'_b)S'_{ab},s'_c]S'M'_S} &=
  \braket{[(s_as_b)S_{ab},s_c]SM_S|[(s'_as'_b)_{13}S'_{ab},s'_c]S'M'_S}
\notag\\&
  =
  \kronecker{s'_as'_cs'_bS'M'_S}{s_as_bs_cSM_S} (-1)^{s_b+s_c+S_{ab}+S'_{ab}} \hat{S}_{ab}\hat{S}'_{ab}\sixj{s_b}{s_a}{S_{ab}}{s_c}{S}{S'_{ab}}
\end{align}
and the isospin part gets an additional factor $\kronecker{\Strng'_a\Strng'_c\Strng'_b}{\Strng_a\Strng_b\Strng_c}$.

The effect on the spatial part is similar: the result of the permutation is a state with the same quantum numbers, but in the coordinate system $\xi'_i$ where the first Jacobi coordinate is defined by particles $a$ and $c$.
Hence,
\begin{align}
\braket{n_\text{cm}l_\text{cm}m_\text{cm},(n_1l_1n_2l_2)LM_L|\op{P}_{bc}|n'_\text{cm}l'_\text{cm}m'_\text{cm},(n'_1l'_1n'_2l'_2)L'M'_L}
&=\kronecker{n'_\text{cm}l'_\text{cm}m'_\text{cm}L'M'_L}{n_\text{cm}l_\text{cm}m_\text{cm}LM_L}\hob{n'_1l'_1}{n'_2l'_2}{n_1l_1}{n_2l_2}{L}{D}
\end{align}
with transformation parameter $D=m_bm_c/(m_a(m_a+m_b+m_c))$.
This relation shows a unique property of the antisymmetrizer in a Jacobi HO basis: the HOB conserves the intrinsic energy quantum number $E=2n_1+l_1+2n_2+l_2$, so the antisymmetrizer is block diagonal not only in $J$ and $T$, but also in $E$ and antisymmetrization can be carried out separately for each (finite-dimensional) block.

To shorten the following formulae, we introduce
\begin{equation}
  \Delta^{a'b'c'}_{abc} = \kronecker{s'_as'_bs'_c}{s_as_bs_c} \kronecker{t'_at'_bt'_c}{t_at_bt_c} \kronecker{\Strng'_a\Strng'_b\Strng'_c}{\Strng_a\Strng_b\Strng_c},
\end{equation}
assume that all particles have spin $s$, and omit the center-of-mass degrees of freedom of which the antisymmetrizer is independent.
Employing \cref{eq:ynn:alpha basis} we get
\begin{align}
  \MoveEqLeft \prescript{}{ab}{\braket{\alpha|\op{P}_{bc}|\alpha'}}_{ab} =
   -\bigl(2+2\kronecker{s_at_a\Strng_a}{s_bt_b\Strng_b}\bigr)^{-1/2}
    \bigl(2+2\kronecker{s'_at'_a\Strng'_a}{s'_bt'_b\Strng'_b}\bigr)^{-1/2}
    \kronecker{\Jcal'\Mcal'}{\Jcal\Mcal} \kronecker{T'M'_T}{TM_T}
    \jhat_1\jhat'_1\jhat_2\jhat'_2
    \hat{S}_{ab}\hat{S}'_{ab}\hat{T}_{ab}\hat{T}'_{ab}
    \sum_{LS} \hat{L}^2 \hat{S}^2
    \ninej{l_1}{S_{ab}}{j_1}{l_2}{s}{j_2}{L}{S}{\Jcal}
    \ninej{l'_1}{S'_{ab}}{j'_1}{l'_2}{s}{j'_2}{L}{S}{\Jcal} \notag\\
  &\times \sixj{s}{s}{S_{ab}}{s}{S}{S'_{ab}} \Biggl(
    \sixj{t_b}{t_a}{T_{ab}}{t_c}{T}{T'_{ab}}
    \hob{n_1l_1}{n_2l_2}{n'_1l'_1}{n'_2l'_2}{L}{D}
    \biggl(
      (-1)^{S_{ab}+S'_{ab}+t_b+t_c+T_{ab}+T'_{ab}} \Delta^{a'c'b'}_{abc}
      + (-1)^{l'_1+S_{ab}+t_a+t_b+2t_c+T_{ab}} \Delta^{b'c'a'}_{abc}\biggr) \notag\\
    &\mindent\phantom{{}={}}
    + \sixj{t_a}{t_b}{T_{ab}}{t_c}{T}{T'_{ab}}
    \hob{n_1l_1}{n_2l_2}{n'_1l'_1}{n'_2l'_2}{L}{D'}
    \biggl(
      (-1)^{l_1+l'_1+2(t_a+t_b+t_c)} \Delta^{c'b'a'}_{abc}
      + (-1)^{l_1+S'_{ab}+2t_a+t_b+t_c+T'_{ab}} \Delta^{c'a'b'}_{abc}\biggr)
  \Biggr) \label{eq:ynn:pab}
\end{align}
with $D'=m_am_c/(m_b(m_a+m_b+m_c))=D[a\leftrightarrow b]$.
The antisymmetrizer is block-diagonal with respect to the quantum numbers $T,M_T,\Jcal,\Mcal,E$ and the orderless set of quantum numbers $\Xcal=\{(s_at_a\Strng_a)$, $(s_bt_b\Strng_b)$, $(s_ct_c\Strng_c)\}$ defining the species of the participant particles.
It is also independent of the projection quantum numbers $\Mcal$ and $M_T$.

The eigenvectors $\vect{c}^{(EJ\Xcal T)}_{i}$ of the matrix representation of $\op{P}_{bc}$ to eigenvalue $-1$ define an orthonormal basis
\begin{equation}
  \ket{Ei\Jcal\Mcal\Xcal TM_T}_a =
    \sum_\alpha
    c^{(EJ\Xcal T)}_{i,\JTred{\alpha}} \ket{\alpha}_{ab}
\end{equation}
spanning the antisymmetric subspace.
The index $i$ enumerates the different vectors emerging from the highly-degenerate eigenvalue problem and carries no physical significance.
The components $c^{(EJ\Xcal T)}_{i,\JTred{\alpha}}$ of the eigenvectors are the coefficents of fractional parentage (CFPs), where $\JTred{\alpha} = \alpha\setminus M_T$.
\end{widetext}

\subsubsection{Transformation to \texorpdfstring{$m$}{m} scheme}
In order to exploit the symmetries of the Hamiltonian and to limit the number of matrix elements that need to be stored in memory during the many-body calculation tractable we store the three-body matrix elements in a $JT$\nobreakdash-coupled scheme.
Hence, we need to express a Slater determinant $\ket{abc}_a$ in terms of antisymmetric Jacobi-HO states $\ket{Ei\Jcal\Mcal\Xcal TM_T}_a$ as follows:
\begin{align}
  \MoveEqLeft\ket{abc}_a = \sqrt{3!}\sum_{J_{ab}T_{ab}}\sum_{JT}\sum_{\alpha}\sum_{n_\text{cm}l_\text{cm}}\sum_{m_\text{cm}}\sum_{i} \notag\\*
  &\times \clebsch{j_a}{m_a}{j_b}{m_b}{J_{ab}}{M_{ab}} \clebsch{J_{ab}}{M_{ab}}{j_c}{m_c}{J}{M} \clebsch{t_a}{\tau_a}{t_b}{\tau_b}{T_{ab}}{\tau_{ab}} \clebsch{T_{ab}}{\tau_{ab}}{t_c}{\tau_c}{T}{M_T} \clebsch{l_\text{cm}}{m_\text{cm}}{\Jcal}{\Mcal}{J}{M} \notag\\
  &\times \Tcoeff{a}{b}{c}{J_{ab}}{T_{ab}}{J}{n}{l}{\alpha} c^{(E\Jcal\Xcal T)}_{i,\JTred{\alpha}} \ket{n_\text{cm}l_\text{cm}m_\text{cm},Ei\Jcal\Mcal\Xcal T M_T}_a.
\end{align}
The sum is over those $\alpha$ with fixed $T$ and $T_{ab}$.
The coefficient
\begin{align}
  \MoveEqLeft \Tcoeff{a}{b}{c}{J_{ab}}{T_{ab}}{J}{n}{l}{\alpha} \notag\\*
    &= \prescript{}{ab}{\braket{(n_\text{cm}l_\text{cm},\alpha)JM|[(\JTred{a}\JTred{b})J_{ab}T_{ab},\JTred{c}]JMTM_T}} \notag\\
    &= \bigl(2+2\kronecker{s_at_a\Strng_a}{s_bt_b\Strng_b}\bigr)^{-1/2} \Tncoeff{a}{b}{c}{J_{ab}}{T_{ab}}{J}{n}{l}{\alpha} \notag\\
    &\hphantom{{}={}}\times\bigl(\Delta^{(abc)_\alpha}_{abc} - (-1)^{l_1+s_a+s_b-S_{ab}+t_a+t_b-T_{ab}} \Delta^{(bac)_\alpha}_{abc}\bigr)
\end{align}
is the overlap of an $ab$-antisymmetric Jacobi-HO basis state and a product state of single-particle HO states coupled to total isospin and angular momentum.
Here, $(abc)_\alpha$ refers to the isospin quantum numbers of the $\alpha$ set.
It can be represented in terms of an overlap between nonantisymmetric states
\begin{align}
  \MoveEqLeft\Tncoeff{a}{b}{c}{J_{ab}}{T_{ab}}{J}{n}{l}{\alpha} =\sum_{L_{ab}} \sum_{N_1L_1} \sum_{LS} \sum_{\Lambda\mathcal{L}} (-1)^{\Lambda+L_{ab}+L+S+J+l_1+l_c} \notag\\
  &\times \jhat_a\jhat_b\jhat_c\jhat_1\jhat_2\hat{L}^2_{ab}\hat{S}_{ab}\hat{J}_{ab}\hat{L}^2\hat{S}^2\hat{\Lambda}^2\hat{\mathcal{L}}^2\hat{\Jcal} \notag\\
  &\times \ninej{l_a}{l_b}{L_{ab}}{s_a}{s_b}{S_{ab}}{j_a}{j_b}{J_{ab}}\ninej{L_{ab}}{l_c}{L}{S_{ab}}{s_c}{S}{J_{ab}}{j_c}{J} \ninej{l_1}{l_2}{\mathcal{L}}{S_{ab}}{s_c}{S}{j_1}{j_2}{\Jcal} \notag\\
  &\times \sixj{l_1}{L_1}{L_{ab}}{l_c}{L}{\Lambda} \sixj{l_\text{cm}}{l_2}{\Lambda}{l_1}{L}{\mathcal{L}} \sixj{l_\text{cm}}{\mathcal{L}}{L}{S}{J}{\Jcal} \notag\\
  &\times \hob{n_al_a}{n_bl_b}{N_1L_1}{n_1l_1}{L_{ab}}{\frac{m_a}{m_b}} \notag\\
  &\times \hob{N_1L_1}{n_cl_c}{n_\text{cm}l_\text{cm}}{n_2l_2}{\Lambda}{\frac{m_a+m_b}{m_c}}.
\end{align}
The $T$ coefficients do not depend on the quantum numbers $M$, $M_T$ and $T$.
In addition to that, we get energy conservation constraints from the HOBs that force $e_a + e_b + e_c = 2n_\text{cm} + l_\text{cm} + E$ and effectively eliminate the sum over $N_1$.

For matrix elements of a scalar operator independent of the center-of-mass degrees of freedom we get
\begin{align}
  \MoveEqLeft \prescript{}{a}{\braket{abc|\op{V}|a'b'c'}}_a = \sum_{J_{ab}J'_{ab}}\sum_{T_{ab}T'_{ab}}\sum_{JT} \clebsch{j_a}{m_a}{j_b}{m_b}{J_{ab}}{M_{ab}} \clebsch{J_{ab}}{M_{ab}}{j_c}{m_c}{J}{M} \clebsch{t_a}{\tau_a}{t_b}{\tau_b}{T_{ab}}{\tau_{ab}} \notag\\
  &\times \clebsch{T_{ab}}{\tau_{ab}}{t_c}{\tau_c}{T}{M_T} \clebsch{j'_a}{m'_a}{j'_b}{m'_b}{J'_{ab}}{M'_{ab}} \clebsch{J'_{ab}}{M'_{ab}}{j'_c}{m'_c}{J}{M} \clebsch{t'_a}{\tau'_a}{t'_b}{\tau'_b}{T'_{ab}}{\tau'_{ab}} \clebsch{T'_{ab}}{\tau'_{ab}}{t'_c}{\tau'_c}{T'}{M'_T} \notag\\
  &\times \Biggl\lbrack 3! \sum_{\alpha\alpha'} \sum_{n_\text{cm}l_\text{cm}} \sum_{ii'} \Tcoeff{a}{b}{c}{J_{ab}}{T_{ab}}{J}{n}{l}{\alpha} \Tcoeff{a'}{b'}{c'}{J'_{ab}}{T'_{ab}}{J}{n}{l}{\alpha'} \notag\\
  &\hphantom{\times\Biggl\lbrack}\times c^{(E\Jcal\Xcal T)}_{i,\JTred{\alpha}} c^{(E'\Jcal\Xcal' T')}_{i',\JTred{\alpha}'} \prescript{}{a}{\braket{Ei\Jcal\Xcal T M_T|\op{V}|E'i'\Jcal\Xcal' T' M'_T}}_a \Biggr\rbrack
\end{align}
and we precompute and store the expression in brackets.
The final decoupling to the $m$-scheme is done on-the-fly during the many-body calculation.

\subsubsection{Embedding of two-body matrix elements}
The evolution in three-body space requires matrix elements of $\Tint$ and $\Ham$ with respect to the antisymmetric Jacobi-HO basis.
For the two-body parts of these operators, we compute
\begin{align}
    \MoveEqLeft \prescript{}{a}{\braket{Ei\Jcal\Xcal T M_T|\op{O}|E'i'\Jcal\Xcal' T' M'_T}}_a \notag\\*
    &= \tfrac{A(A-1)}{2} \sum_{\alpha,\alpha'} c^{(E\Jcal\Xcal T)}_{i,\JTred{\alpha}} c^{(E'\Jcal\Xcal' T')}_{i',\JTred{\alpha}'} \prescript{}{ab}{\braket{\alpha|\op{o}_{ab}|\alpha'}}_{ab}
\end{align}
where $\op{O}$ denotes a general two-body operator, embedded into three-body space.
The operator $\op{o}_{ab}$ acts only on the first two particles.
The $ab$-antisymmetric Jacobi-HO basis has the same quantum numbers and coupling scheme as a $JT$\nobreakdash-coupled two-body basis.
Thus, assuming a scalar-isoscalar operator, the three-body matrix elements are calculated from the two-body ones by
\begin{align}
  \prescript{}{ab}{\braket{\alpha|\op{o}_{ab}|\alpha'}}_{ab} &=\prescript{}{a}{\bra{[n_1l_1,(s_as_b)S_{ab}]j_1,(\Strng_at_a\Strng_bt_b)T_{ab}}} \notag\\*
  &\phantom{{}={}}\times\op{o}\ket{[n'_1l'_1,(s'_as'_b)S'_{ab}]j_1,(\Strng'_at'_a\Strng'_bt'_b)T_{ab}}_{a} \notag\\
  &\phantom{{}={}}\times\kronecker{j'_1n'_2l'_2s'_cj'_2\Strng'_ct'_cT'_{ab}}{j_1n_2l_2s_cj_2\Strng_ct_cT_{ab}}.
\end{align}
For nonscalar operators, one needs to decouple the second Jacobi coordinate and the dependence on the projection quantum numbers introduces three-body matrix elements coupling different $\Jcal$ (or $T$).

\subsubsection{Numerical solution}
If we neglect isospin breaking, the Hamiltonian of the three-body system decomposes into blocks with good $\Strng \Jcal^\Pi T$ with intrinsic parity $\Pi$ and the SRG evolution acts on each block separately.
This makes a straight-forward solution of the flow equation feasible.
We solve the matrix differential equation \cref{eq:ynn:flow equation} with standard numerical methods, calculating the derivative by computing the double matrix commutator.
Since all matrices and intermediates are (anti-) hermitian
\begin{equation}
  A^\dag = \sigma_A A,
\end{equation}
with $\sigma_A=+1$ ($\sigma_A=-1$), the matrix commutator is
\begin{equation}
  \comm{A}{B} = AB-BA = AB - (A^\dag B^\dag)^\dag = AB - \sigma_A\sigma_B (AB)^\dag
\end{equation}
and we can compute it by a single matrix multiplication followed by an addition or subtraction of the adjoint of the result.

\subsubsection{Subtraction}
After the evolution the three-body matrix elements contain a mixture of intrinsic kinetic energy, two- and induced three-body interactions.
The induced three-body terms need to be separated because two- and three-body interactions scale differently in many-body calculations, i.e., one cannot simply embed the three-body matrix elements of the Hamiltonian into $A$\nobreakdash-body space.
We achieve the separation by subtracting the intrinsic kinetic energy and a two-body interaction evolved in two-body space to the same flow parameter.

Here, one has to carefully consider the truncations of the two- and three-body SRG model spaces: to subtract truncation artifacts, one has to, in priciple, truncate two-body matrix elements identically in both model spaces.
The three-body space is truncated by $E \leq E_{3,\text{max}}$, so the maximum relative energy of the first two particles (the first Jacobi coordinate) depends on the energy of the spectator particle.
There is hence no single truncation $E_{2,\text{max}}$ for the two-body evolution to describe this truncation.
So, to capture the main effect and to minimize truncation artifacts in the first place, we set $E_{2,\text{max}} = E_{3,\text{max}}$ and choose sufficiently large SRG model spaces by setting $E_{3,\text{max}}$ and the basis frequency $\Omega$ to sufficiently high values.

A consequence of this discussion is that the optimal basis frequencies for the SRG evolution and for the many-body calculation are in general different.
Before converting the matrix elements to single-particle coordinates we, therefore, change the basis frequency from the former to the latter.
The method is the same as described in Ref.\ \cite{Roth2014} and the overlap of two states with different basis frequencies $\Omega$ and $\Omega'$ is given by
\begin{multline}
  \shoveleft{\prescript{\Omega}{a}{\braket{Ei\Jcal\Xcal T M_T|E'i'\Jcal'\Xcal' T' M'_T}}^{\Omega'}_a } \\
   = \sum_{\alpha\alpha'} c^{(E\Jcal\Xcal T)}_{i,\JTred{\alpha}} c^{(E'\Jcal'\Xcal' T')}_{i',\JTred{\alpha}'} \prescript{\Omega}{ab}{\braket{\alpha|\alpha'}}^{\Omega'}_{ab}
\end{multline}
with
\begin{align}
  \MoveEqLeft\prescript{\Omega}{ab}{\braket{\alpha|\alpha'}}^{\Omega'}_{ab} = \notag\\
   &\kronecker{\Jcal \Mcal S_{ab} T_{ab} T M_T}{\Jcal' \Mcal' S'_{ab} T'_{ab} T' M'_T}\kronecker{l_1 j_1 l_2 j_2}{l'_1 j'_1 l'_2 j'_2} \bigl(1+\kronecker{s_at_a\Strng_a}{s_bt_b\Strng_b}\bigr)^{-1/2} \bigl(1+\kronecker{s'_at'_a\Strng'_a}{s'_bt'_b\Strng'_b}\bigr)^{-1/2} \notag\\
   &\times (\Delta_{abc}^{a'b'c'} + (-1)^{l_1 + S_{ab} + t_a + t_b - T_{ab}} \Delta_{abc}^{b'a'c'}) \notag\\
   &\times \int \mathrm{d}r_1 r_1^2 R_{n_1l_1}(r_1,b(\mu_1,\Omega)) R_{n'_1l_1}(r_1,b(\mu_1,\Omega')) \notag\\
   &\times \int \mathrm{d}r_2 r_2^2 R_{n_2l_2}(r_2,b(\mu_2,\Omega)) R_{n'_2l_2}(r_2,b(\mu_2,\Omega')).
\end{align}
The functions $R_{nl}(r,b)$ are HO wave functions with oscillator lengths $b(\mu,\Omega)=(\mu\Omega)^{-1/2}$, and $\mu_1 = (m_a^{-1}+m_b^{-1})^{-1}, \mu_2 = ((m_a+m_b)^{-1}+m_c^{-1})^{-1}$ are the reduced masses corresponding to the Jacobi coordinates.
The integrals can be transformed so they only depend on the ratio $\Omega'/\Omega$ of the basis frequencies.
Analytical expressions for the overlaps are given in \cref{app:hooverlap}.

As a final remark we note that the antisymmetrization and embedding procedures presented here actually comprise a Jacobi-coordinate formulation of the NCSM (J-NCSM) for the three-body system.
This formulation is very economical because it makes extensive use of the symmetries of the system.
An extension to larger particle numbers is feasible, and has been employed for nuclear \cite{Barrett2013,Liebig2016} and hypernuclear \cite{Wirth2018a} systems, to a point where antisymmetrization becomes too cumbersome.

\section{\label{sec:ncsm}No-core shell model}
The SRG-evolved Hamiltonian can be used in any many-body method, either directly or through additional approximations for the three-body force like the normal-ordered two-body approximation \cite{Roth2012}.
One of the conceptually simplest many-body methods able to include the full three-body part of the Hamiltonian is the no-core shell model (NCSM).
In the NCSM, a matrix representation of the Hamiltonian is computed in a model space spanned by Slater determinants of HO single-particle states.
The model space is truncated by limiting the number of HO excitation quanta to a maximum value of $\Nmax$.
Since the basis states are Slater determinants, antisymmetrization is trivial and computation of many-body matrix elements using Slater-Condon rules is simple.
Also, working with single-particle states allows for an exact treatment of isospin-breaking mass differences, the Coulomb interaction and charge-symmetry-breaking parts of the baryon-baryon interactions.

Due to the $\Lambda$-$\Sigma$ conversion present in the YN interaction, the model space for hypernuclei contains determinants with different numbers of protons, neutrons and hyperons.
Only the total number of particles, total charge and strangeness are conserved.
This increases the size of the model spaces significantly and, in combination with the rapid growth with total particle number and $\Nmax$, limits the range of tractable hypernuclei.
This limitation is mitigated by the introduction of an importance-truncation scheme \cite{Roth2009}, where, starting from a reference state from a small model space, the importance of each basis state for the description of the target state is estimated perturbatively.
Only those states whose importance exceeds a given threshold are included in the (importance-truncated) model space.
The effect of the neglected states on observables is taken into account by computing them for multiple values of the threshold and subsequently extrapolating to vanishing threshold.

Starting from a small-$\Nmax$ calculation, one can thus build an iterative scheme, where the eigenstates computed in each $\Nmax$ step are used as reference states for constructing the model space for the next step.
During this procedure, all basis states are reassessed and the model space dynamically adapts to the structure of the targeted states.

\section{\label{sec:results}Results}

\subsection{Interaction dependence of hypernuclear states}
\begin{figure}
  \centering
  \includegraphics{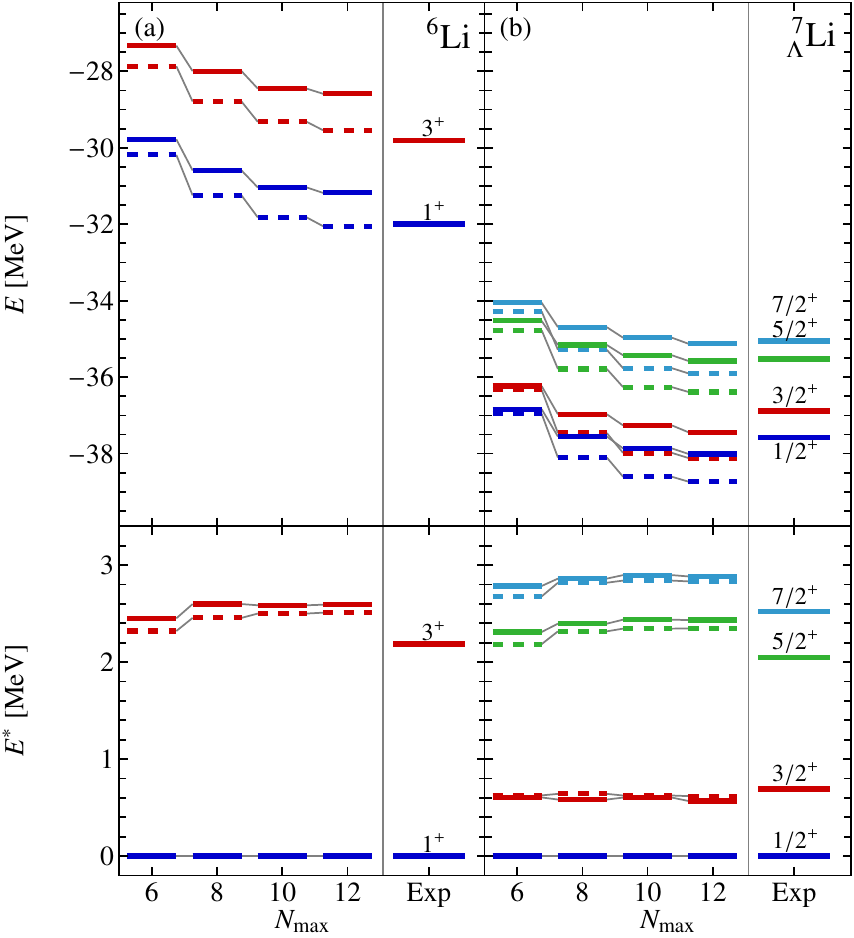}
  \caption{\label{fig:lli7}\coloronline Absolute and excitation energies of low-lying states of (a) \isotope[6]{Li} and (b) its daughter hypernucleus \isotope[7][\Lambda]{Li} for the \EMNfourNL{} (solid lines) and the \EMthreeL{} (dashed lines) nucleonic Hamiltonians.
  See text for details.
  }
\end{figure}
\begin{figure}
  \centering
  \includegraphics{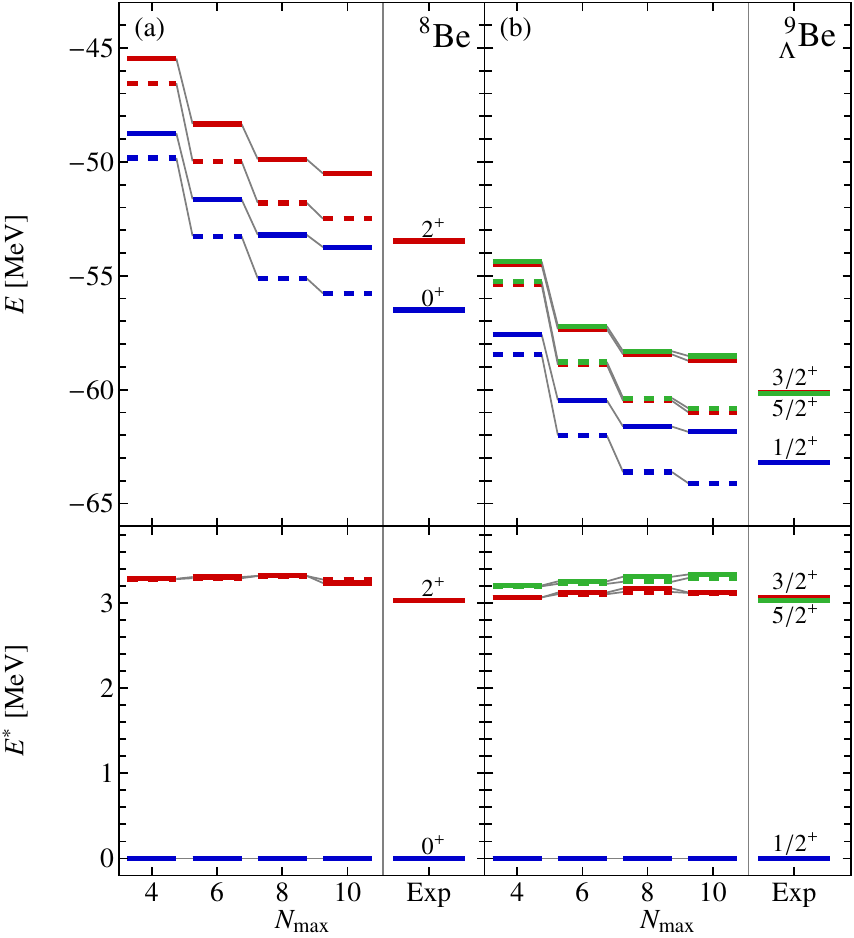}
  \caption{\label{fig:lbe9}\coloronline Same as \cref{fig:lli7}, but for (a) \isotope[8]{Be} and (b) \isotope[9][\Lambda]{Be}.
  }
\end{figure}
\begin{figure}
  \centering
  \includegraphics{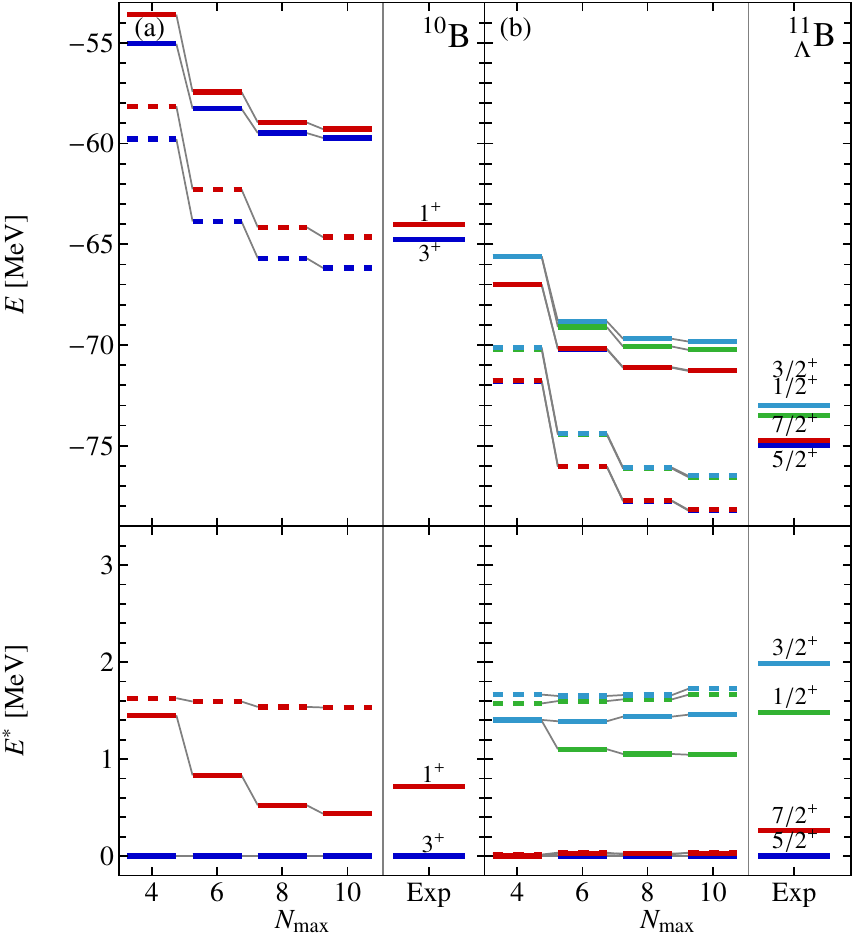}
  \caption{\label{fig:lb11}\coloronline Same as \cref{fig:lli7}, but for (a) \isotope[10]{B} and (b) \isotope[11][\Lambda]{B}.
  }
\end{figure}
\begin{figure}
  \centering
  \includegraphics{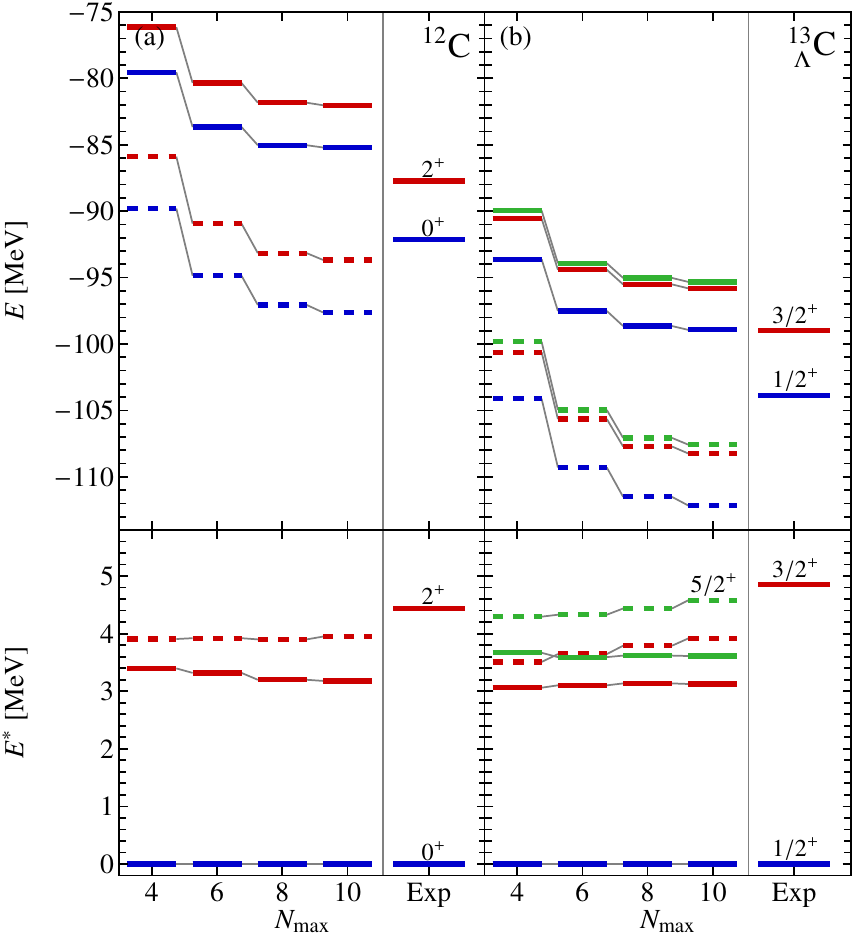}
  \caption{\label{fig:lc13}\coloronline Same as \cref{fig:lli7}, but for (a) \isotope[12]{C} and (b) \isotope[13][\Lambda]{C}.
  }
\end{figure}

Our previous calculations did not explore the effect of the nucleonic Hamiltonian on hypernuclear observables.
Instead, we only used a single Hamiltonian that provides a good reproduction of experimental energy levels in the $s$ and $p$ shells.
This Hamiltonian, however, has certain deficiencies.
First, it predicts nuclear radii that are too small by approximately \SI{20}{\percent}.
Second, there was an error in the formula for determining the three-nucleon low-energy constant $c_D$ from the triton beta-decay half life \cite{Marcucci2018,Gazit2019}.
Recently, new families of interactions became available that use the correct formula and predict larger radii.

Larger nuclear radii imply lower nucleon densities in the interior of the nucleus.
Since the YN interaction essentially couples the hyperon to the nucleon density, we expect that the hyperon will experience less attraction, leading to a lower hyperon separation energy.

In the following, we will use three different Hamiltonians.
The first is the previously used one, consisting of a two-nucleon interaction at next-to-next-to-next-to-leading order (N\textsuperscript3LO) by \citet{Entem2003} and a three-nucleon interaction at N\textsuperscript2LO with local regulator \cite{Navratil2007,Gazit2009}.
We will refer to this Hamiltonian as \EMthreeL.
for the second Hamiltonian, called \EMthreeNL, we use a nonlocal regulator for the three-nucleon interaction with the corrected $c_D$ value \cite{Gazit2019}.
Finally, we consider a Hamiltonian that is built from the recent N\textsuperscript4LO two-nucleon interaction by \citet*{Entem2017} and also uses a nonlocal three-nucleon interaction. In what follows, we call this Hamiltonian \EMNfourNL.
All Hamiltonians use a regulator cutoff of $\Lambda_N=\SI{500}{\MeVc}$ in the two- and three-body sector.
For hypernuclei, we combine them with a LO hyperon-nucleon interaction \cite{Polinder2006} with a cutoff of $\Lambda_Y=\SI{700}{\MeVc}$.
The calculations are carried out with an oscillator frequency of $\hbar\Omega=\SI{20}{\MeV}$, which is close to the variational minimum.
For the \EMthreeL{} Hamiltonian, we use an SRG flow parameter of $\alpha=\SI{0.08}{\fm\tothe4}$ as in our previous publications.
The other Hamiltonians are evolved to $\alpha=\SI{0.12}{\fm\tothe4}$ wich provides faster convergence and an improved description of the \isotope[4]{He} ground-state energy. A detailed discussion of these Hamiltonians and applications to non-strange nuclei is presented elsewhere \cite{Huether2019}.

To investigate the effect of the nucleonic Hamiltonian on the properties of hypernuclei, we consider the hypernuclei \isotope[7][\Lambda]{Li}, \isotope[9][\Lambda]{Be}, \isotope[11][\Lambda]{B}, and \isotope[13][\Lambda]{C}, as well as their nonstrange parent nuclei.
Here, we compare the \EMthreeL{} and the \EMNfourNL{} Hamiltonians in order to assess the effect of switching to the new generation of chiral Hamiltonians.

\Cref{fig:lli7} shows the absolute and excitation energies of \isotope[6]{Li} and \isotope[7][\Lambda]{Li}.
In the absolute energies for the parent nucleus, we see that the \EMNfourNL{} Hamiltonian provides approx.\ \SI{1}{\MeV} less binding for both the $1^+$ ground state and the first excited $3^+$ state.
This feature carries over to the hypernucleus and cancels some of the overbinding generated by the YN interaction, bringing the calculated absolute energies closer to the experimental ones.
Excitation energies of the parent and the hypernucleus show very little variation between the two Hamiltonians.
Only the $3^+$ excitation energy changes by approx.\ \SI{100}{\keV}, which is reflected in the excitation energies of the $5/2^+,7/2^+$ doublet in the hypernucleus.
The $3/2^+$ excitation energy shows almost no variation at all.

The situation for \isotope[8]{Be} and \isotope[9][\Lambda]{Be} (cf.\ \cref{fig:lbe9}) is very similar to \isotope[7][\Lambda]{Li}, just that the \EMNfourNL{} Hamiltonian now provides \SI{2}{\MeV} less binding energy.

The picture changes for \isotope[11][\Lambda]{B}, shown in \cref{fig:lb11}.
The excited $1^+$ state in \isotope[10]{B} already behaves differently for the two Hamiltonians. For the \EMNfourNL{} the excitation energy converges slowly to a value approximately half of the experimental energy, while the \EMthreeL{} predicts a much higher excitation energy that is almost independent of $\Nmax$.
Adding a hyperon, the $3^+$ ground state splits into a near-degenerate doublet with a splitting of less than \SI{50}{\keV}.
While the excited state doublet is situated around the same energy as the $1^+$ parent state with small doublet splitting for the \EMthreeL{} Hamiltonian, it shifts by \SI{0.8}{\MeV} to higher excitation energies with the \EMNfourNL{} Hamiltonian.
Also, the excited doublet splitting is larger and more in line with experimental data.
Both Hamiltonians seem to predict very different structures for the first excited state \isotope[10]{B}, leading to the different behavior of the hypernuclear states.

In \isotope[12]{C} (cf.\ \cref{fig:lc13}), the binding energy difference increases to more than \SI{10}{\MeV}.
Additionally, the excitation energy of the $2^+$ is almost \SI{1}{\MeV} lower for the \EMNfourNL{}, increasing the difference to the experimental value.
The lower excitation energy translates to the excited-state doublet in the hypernucleus, where we also notice slightly different doublet spacings and a different convergence behavior.

\begin{table}
  \begin{tabular}{%
    r%
    S[table-format=+3.2(2)]%
    S[table-format=+2.2(2)]%
    S[table-format=+3.2(2)]%
  }
    \toprule
    {} & {\EMthreeL} & {\EMNfourNL} & {Exp. \cite{Wang2012,Davis2005}} \\
    \midrule
    \isotope[6]{Li}          & -32.36(4)  & -31.44(8)  & -31.99      \\[0.2ex]
    \isotope[7][\Lambda]{Li} & -39.25(4)  & -38.10(1)  & -37.57(3)   \\[0.2ex]
    \isotope[8]{Be}          & -56.24(29) & -54.13(17) & -56.50      \\[0.2ex]
    \isotope[9][\Lambda]{Be} & -64.7(4)   & -62.25(29) & -63.21(4)   \\[0.2ex]
    \isotope[10]{B}          & -67.0(6)   & -60.20(34) & -64.75      \\[0.2ex]
    \isotope[11][\Lambda]{B} & -78.8(6)   & -71.58(34) & -74.99(5)   \\[0.2ex]
    \isotope[12]{C}          & -98.7(8)   & -85.8(5)   & -92.16      \\[0.2ex]
    \isotope[13][\Lambda]{C} & -113.1(8)  & -99.36(38) & -103.85(12) \\
    \bottomrule
  \end{tabular}

  \caption{\label{tab:e0 extrapolated}%
  Extrapolated ground-state energies (in MeV) for the \EMthreeL{} and \EMNfourNL{} Hamiltonians, compared to experimental data.
  The extrapolation prescription is the same as in \cite{Wirth2018}.
  }
\end{table}

\begin{table}
  \begin{tabular}{%
    r%
    S[table-format=2.2(2)]%
    S[table-format=2.2(2)]%
    S[table-format=2.2(2)]%
  }
    \toprule
    {} & {\EMthreeL} & {\EMNfourNL} & {Exp. \cite{Davis2005}} \\
    \midrule
    \isotope[7][\Lambda]{Li} & 6.89(6)  & 6.66(8)  & 5.58(3)   \\[0.2ex]
    \isotope[9][\Lambda]{Be} & 8.5(5)   & 8.12(34) & 6.71(4)   \\[0.2ex]
    \isotope[11][\Lambda]{B} & 11.9(8)  & 11.4(5)  & 10.24(5)  \\[0.2ex]
    \isotope[13][\Lambda]{C} & 14.5(11) & 13.5(6)  & 11.69(12) \\
    \bottomrule
  \end{tabular}
  \caption{\label{tab:blambda}%
  Calculated hyperon separation energies (in MeV) for the \EMthreeL{} and \EMNfourNL{} Hamiltonians, compared to experimental data.
  }
\end{table}

If we look at the $N_{\text{max}}$-extrapolated ground-state energies tabulated in \cref{tab:e0 extrapolated}, we notice that the results for the parent nuclei bracket the experimental value in all cases except for \isotope[8]{Be}, which is a narrow resonance above the $2\alpha$ threshold and is expected to get significant continuum corrections, which we do not include in our calculations.
Due to the cancellation of under- and overbinding in the nucleonic and hyperonic Hamiltonian, respectively, the \EMNfourNL{} results for the hypernuclei are closer to the experimental values.
Surprisingly, and contrary to our initial expectations, the hyperon separation energies (cf.\ \cref{tab:blambda}) are not changed dramatically by changing the nucleonic Hamiltonian.
While the central values are lowered systematically for the \EMNfourNL{} Hamiltonian, the results for both Hamiltonians are compatible within extrapolation uncertainties.

\begin{figure*}
  \centering
  \includegraphics{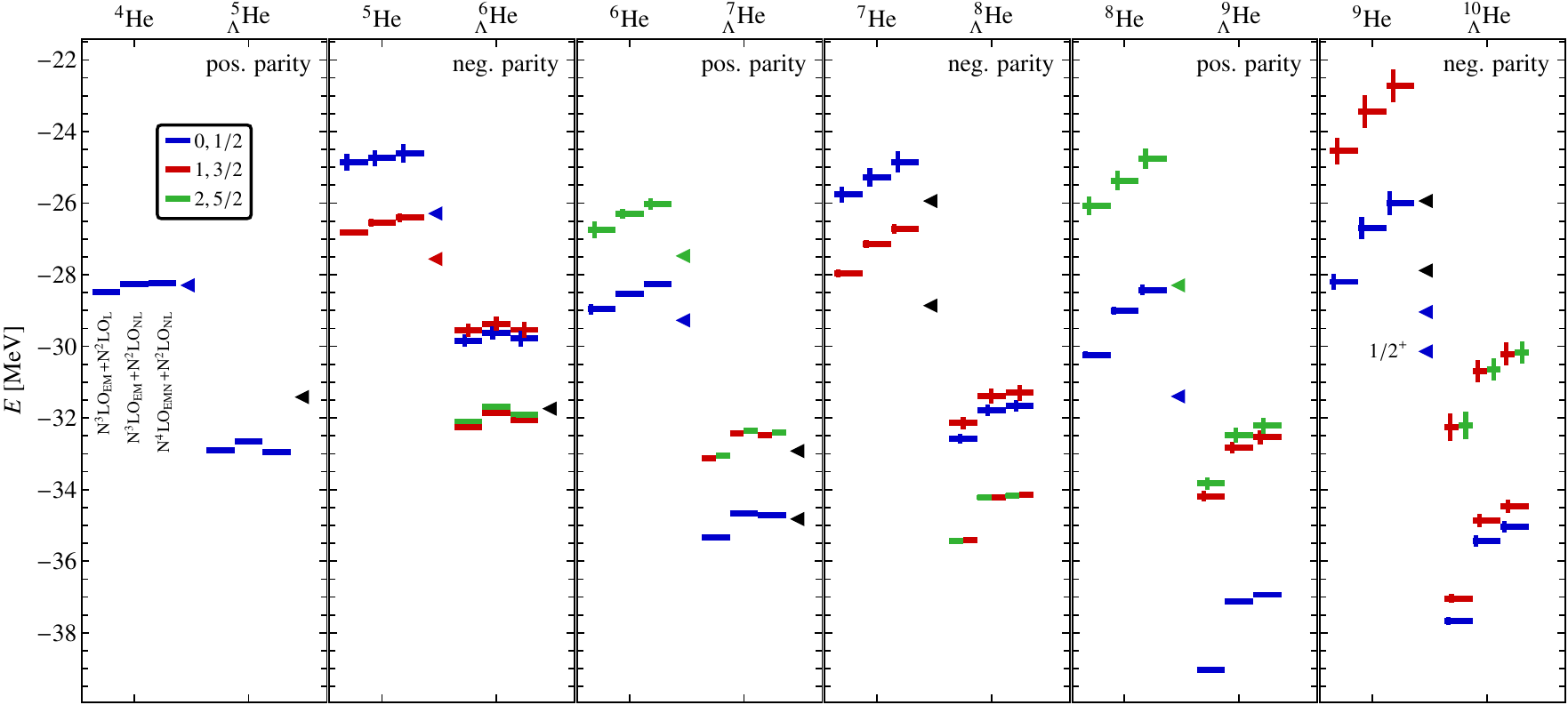}
  \caption{\label{fig:he chain}%
    \coloronline%
    Extrapolated absolute energies of low-lying natural-parity states in the hyper-helium chain and their parent nuclei for the \EMthreeL{}, \EMthreeNL{}, and \EMNfourNL{} Hamiltonian.
    Experimental values from \cite{Wang2012,Tilley2002,Tilley2004,Davis2005,Gogami2016a}.
  }
\end{figure*}
\begin{figure*}
  \centering
  \includegraphics{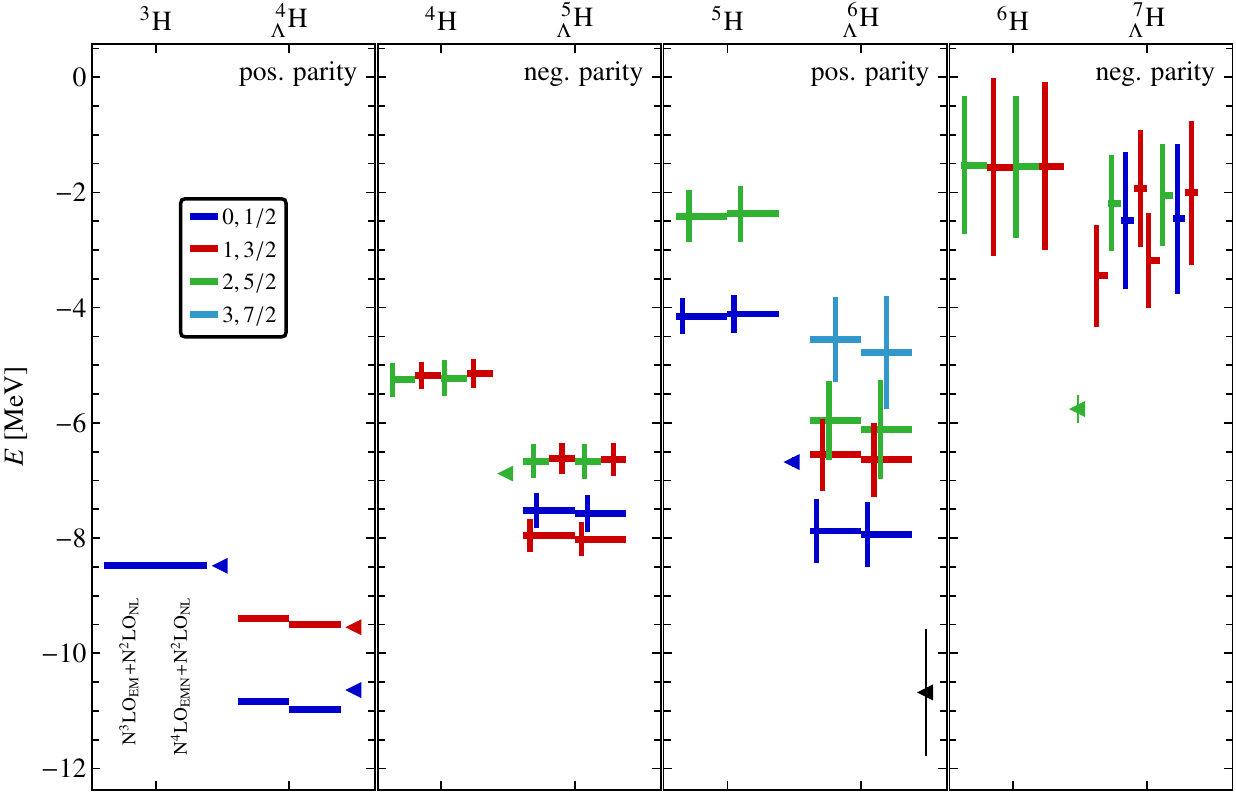}
  \caption{\label{fig:h chain}%
    \coloronline%
    Extrapolated absolute energies of low-lying natural-parity states in the hyper-hydrogen chain and their parent nuclei for the \EMthreeNL{} and \EMNfourNL{} Hamiltonian.
    Experimental data are marked by triangles where available \cite{Wang2012,Schulz2016,Agnello2012}. The \isotope[3]{H} binding energy was used to fit the NNN interaction low-energy constants.
  }
\end{figure*}

\subsection{Effect of the Nucleon-nucleon interaction in neutron-rich hypernuclei}
The nucleonic Hamiltonian also plays an important role in determining the structure of light neutron-rich hypernuclei.
These systems probe the isospin dependence of the two- and three-nucleon interaction.
To assess the uncertainty stemming from the two-nucleon interaction, we consider the hyper-helium and -hydrogen chains and compare the \EMNfourNL{} and \EMthreeNL{} Hamiltonians.

First, we consider hypernuclei from the helium chain, where the parent nuclei are either bound or relatively close to the threshold.
The extrapolated absolute energies are shown in \cref{fig:he chain} for \isotope[4,\dotsc,9]{He} and their daughter hypernuclei.
For reference, we also show the results using the \EMthreeL{} Hamiltonian, which we discussed in a previous work \cite{Wirth2018}.

In the energies of the nucleonic states, we see a systematic trend towards less binding when moving through the three Hamiltonians.
The energy difference increases with neutron number, and the largest jump is caused by changing the regulator of the three-nucleon force from local to nonlocal.
Thus, it seems that the three-nucleon interaction is the leading cause for the differences between the predictions by the three Hamiltonians.

In the results for the hypernuclei, we clearly see a difference between the hyperon separation energies predicted by the \EMthreeNL{} and \EMNfourNL{} Hamiltonians.
The \EMNfourNL{} Hamiltonian predicts higher ground-state energies for all parent nuclei, but the hypernuclear ground states up to \isotope[7][\Lambda]{He} are at lower energies than for the \EMthreeNL{} Hamiltonian.
In the more neutron-rich isotopes, they are much closer than the ground-state energies of the parent nuclei.
Hence, the \EMNfourNL{} Hamiltonian produces noticeably larger hyperon separation energies.
In line with the findings of the previous section, these separation energies are close to the ones predicted by the \EMthreeL{} Hamiltonian.

This effect is a consequence of a cancellation between the two- and three-nucleon interactions.
When switching from the \EMthreeL{} to the \EMthreeNL{} Hamiltonian, the hyperon separation energies are slightly reduced.
The increase caused by changing the two-body interaction approximately cancels the reduction.
For example, the separation energy in \isotope[7][\Lambda]{He} changes from \SI{6.36(16)}{\MeV} to \SI{6.13(8)}{\MeV} and then back to \SI{6.46(8)}{\MeV}.

We also notice the fortuitous cancellation of under- and overbinding generated by the nucleonic and hyperonic Hamiltonians, which brings the calculations very close to the experimental values.
This reproduction of the experimental values might be advantageous for predicting other observables in these hypernuclei.

Regarding the stability of the hypernuclei, the findings are the same as in \cite{Wirth2018}, where we used the old N\textsuperscript3LO Hamiltonian:
The \isotope[5][\Lambda]{He} hypernucleus is overbound---a known issue for YN interactions that reproduce the $A=4$ binding energies.
In consequence, \isotope[6][\Lambda]{He} is predicted to be unstable.
The daughter hypernuclei of the particle-stable parents \isotope[6,8]{He} are also predicted to be stable, while the ground-state doublet of \isotope[8][\Lambda]{He} is put slightly above the $\isotope[7][\Lambda]{He} + n$ threshold.

In the last part of this survey, we consider systems with extreme neutron-proton asymmetries: the neutron-rich hyper-hydrogen isotopes.
The extrapolated energies of low-lying natural-parity states in \isotope[3,\dotsc,6]{H} and their daughter hypernuclei are shown in \cref{fig:h chain}.
Considering \isotope[3]{H}, whose ground-state energy is used in fitting the parameters of the three-body force, and \isotope[4][\Lambda]{H}, we see some slight differences in the $0^+$ and $1^+$ energies.
Compared to the \EMthreeNL, the \EMNfourNL{} Hamiltonian increases the hyperon separation energy on the scale of \SI{100}{\keV} without visibly affecting the excitation energy of the $1^+$.
The heavier hydrogen nuclei are predicted to be well inside the \isotope[3]{H} plus \text{neutrons} continuum by both Hamiltonians, which both yield values that are compatible within extrapolation uncertainties.
Compared to experimental data, the absolute energies are increasingly underbound with the number of neutrons.
Since we calculate these continuum states in a bound-state approximation, the underbinding is expected to some extent.
In \isotope[4]{H}, we find that the ground state and first excited state form a closely-spaced $2^-,1^-$ doublet, as is observed in experiment.
A similar structure is predicted in \isotope[6]{H}, albeit with much larger extrapolation uncertainties.
This is caused by slow convergence of the absolute energies: the ground-state energy calculated at $N_\text{max}=12$ is still positive.
The two lowest states in \isotope[5]{H} show a larger separation of almost \SI{2}{\MeV}.

The small separation of the \isotope[4]{H} states translates into a peculiar structure of the \isotope[5][\Lambda]{H} spectrum:
the doublets generated by the two states overlap, thus the two lowest hypernuclear states do not form a doublet, but are the lower states of two different doublets, one generated by the $2^-$ and the other generated by the $1^-$ state.

The hypernucleus \isotope[6][\Lambda]{H} is especially interesting because it has been observed experimentally \cite{Agnello2012}.
While our results are clearly underbound compared to the experimental ground-state energy, the hyperon separation energy amounts to approximately \SI{3.8}{\MeV}.
This value is well within the uncertainty of the experimental value of $B_\Lambda=\SI{4.0(11)}{\MeV}$.

In \isotope[7][\Lambda]{H}, we see a closely spaced group of four states generated by the two lowest states in \isotope[6]{H}.
The structure seems to be similar to the one found in \isotope[5][\Lambda]{H}, but large extrapolation uncertainties hinder further interpretation.
The small hyperon separation energy of this hypernucleus is striking: judging from the extrapolated ground-state energies it is half of the hyperon separation energy of \isotope[6][\Lambda]{H}.
This is contrary to the observation that the hyperon separation energy smoothly increases as one adds nucleons to the system.

\section{Conclusions}
In order to get reliable predictions, it is crucial to include induced YNN terms in any calculation that uses SRG-transformed interactions to compute properties of systems containing hyperons.
We present in detail a procedure for obtaining these YNN terms induced during the SRG evolution of a hypernuclear Hamiltonian.
The procedure consists of embedding the NN and YN interactions into a three-body Jacobi HO basis, evolving the Hamiltonian to generate the three-body terms, and isolating them by subtracting the two-body interaction evolved in two-body space from the resulting Hamiltonian.
All steps can be carried out efficiently on individual blocks of the Hamiltonian, because the basis manifestly conserves many of its symmetries.
Finally, we describe the conversion to an efficient $JT$-coupled storage scheme from which $m$-scheme matrix elements can be recovered with little effort.

In the second part of this work, we apply the IT-NCSM to symmetric and neutron-rich hypernuclei, focussing on the effect of the nucleonic Hamiltonian on low-lying hypernuclear states.
First, we compare the tried-and-tested \EMthreeL{} Hamiltonian against the state-of-the-art \EMNfourNL{} Hamiltonian.
A key difference between these two Hamiltonians is that, due to the use of a nonlocal three-body interaction, the \EMNfourNL{} predicts significantly larger radii much closer to the experimental values than those predicted by the \EMthreeL.
As a consequence, the central nucleon density gets reduced and na\"ively leads to a lower hyperon separation energy.
Contrary to this expectation, we find only a very small difference between the separation energies calculated with both Hamiltonians.
However, some hypernuclear states like the excited-state doublet in \isotope[11][\Lambda]{B} are significantly affected by the choice of nuclear Hamiltonian.

Next, we additionally consider a third Hamiltonian, \EMthreeNL, that employs the same two-nucleon interaction as the \EMthreeL{} but substitutes the local three-nucleon interaction by a nonlocal one, in the chain of neutron-rich helium hypernuclei.
We see that the smallness of the difference is caused by a systematic cancellation between a decrease of the separation energy due to the nonlocal three-nucleon interaction and an increase caused by the two-body interaction of the \EMNfourNL{} Hamiltonian.

We finish with a survey of the hyper-hydrogen chain.
Here, we find that the nucleonic parents beyond \isotope[3]{H} are already strongly underbound compared to experiment, and, in consequence, the hypernuclei are all predicted well inside the \isotope[4][\Lambda]{H} plus neutrons continuum.
Overall, large extrapolation uncertainties make an interpretation of the results difficult.

In conclusion, the ability to use SRG-transformed Hamiltonians with induced YNN terms paves the way for a multitude of new developments in the \emph{ab initio} description of strange nuclear systems like the application of perturbation theory or efficient many-body methods for medium-mass systems where the slow rate of convergence of the bare interaction has been an obstacle.

\begin{acknowledgments}
We gratefully acknowledge support by the BMBF through contract 05P18RDFN1 (NuSTAR.DA), the Deutsche Forschungsgemeinschaft (DFG, German Research Foundation) Projektnummer 279384907 (SFB 1245), and the Helmholtz International Center for FAIR.
Calculations for this research were conducted on the LICHTENBERG high-performance computer of TU Darmstadt, on the supercomputer JURECA \cite{JURECA} at Forschungszentrum Jülich, and on the cluster of the Institute for Cyber-Enabled Research at Michigan State University.
\end{acknowledgments}

\appendix

\section{\label{app:hooverlap}Overlap of HO wave functions}
The frequency-conversion step requires the knowledge of overlaps
\begin{spreadlines}{4pt}
\begin{align}
  \MoveEqLeft\braket{\phi_{nlm}(b)|\phi_{n'l'm'}(b')}\notag\\
    &=\int_0^{\infty}\mathrm{d}{r}r^2 \int \mathrm{d}{\Omega}\, R_{nl}(r,b) R_{n'l'}(r,b') Y^*_{lm}(\Omega) Y_{l'm'}(\Omega) \notag\\
    &= \delta_{ll'}\delta_{mm'} \int_0^{\infty}\mathrm{d}{r} r^2 R_{nl}(r,b) R_{n'l}(r,b') \\
    &\equiv \delta_{ll'}\delta_{mm'} I_l(nb,n'b')
\end{align}
\end{spreadlines}
between HO states with different oscillator lengths $b$ and $b'$.
The radial integral can be decomposed further because
\begin{align}
  \MoveEqLeft r R_{nl}(r,b) \notag\\
    &= \sqrt{\frac{2\Gamma(n+1)}{b\Gamma(n+l+3/2)}} \exp\bigl(-\tfrac12 \bigl(\tfrac{r}{b}\bigr)^2\bigr) \bigl(\tfrac{r}{b}\bigr)^{l+1} L_n^{(l+1/2)} \bigl(\bigl(\tfrac{r}{b}\bigr)^2\bigr) \notag\\
    &\equiv \mathcal{N}_{nl} b^{-1/2} f_{nl}\bigl(\tfrac{r}{b}\bigr)
\end{align}
factorizes into a normalization factor $\mathcal{N}_{nl}$ and a function $f_{nl}$ that depends only on the ratio $r/b$.
We thus have
\begin{align}
  I_l(nb,n'b') &= \int_0^{\infty}\mathrm{d}{r} r^2 R_{nl}(r,b) R_{n'l}(r,b') \notag\\
    &= \mathcal{N}_{nl}\mathcal{N}_{n'l} (bb')^{-1/2}\int_0^{\infty}\mathrm{d}{r}\, f_{nl}\bigl(\tfrac{r}{b}\bigr) f_{n'l}\bigl(\tfrac{r}{b'}\bigr)
\end{align}
and by substituting $\rho = r/b$ we can make the integral depend only on the ratio $a = b/b'$ of the oscillator parameters:
\begin{align}
  \MoveEqLeft I_l(nb,n'b') = \mathcal{N}_{nl}\mathcal{N}_{n'l} a^{l+3/2} \int_0^{\infty}\mathrm{d}{\rho} \exp\bigl(-\tfrac12 (1+a^2)\rho^2\bigr)
   \notag\\&\times
   \rho^{2l+2} L_n^{(l+1/2)}(\rho^2) L_{n'}^{(l+1/2)}(a^2\rho^2).
\end{align}
A further substitution $x = 1/2 (1+a^2)\rho^2$ together with the multiplication theorem \cite[eq.~18.18.12]{NIST:DLMF} for associated Laguerre polynomials enables us to carry out the integral, yielding a sum over a finite number of terms:
\begin{align}
  \MoveEqLeft I_l(nb,n'b') \notag\\*&
  = (-1)^n \biggl(\frac{2}{a+a^{-1}}\biggr)^{l+3/2} \bigl(1+a^2\bigr)^{-(n+n')} \sum_{k=0}^{n_<} \frac{(-1)^k}{k!}
    \notag\\&\phantom{{}={}}\times
  \frac{\sqrt{n!n'!}}{(n-k)!(n'-k)!} \frac{\sqrt{\Gamma(n+l+3/2)\Gamma(n'+l+3/2)}}{\Gamma(k+l+3/2)}
    \notag\\&\phantom{{}={}}\times
  (2a)^{2k} (1-a^2)^{n+n'-2k} \\
  &= (-1)^n \biggl(\frac{2}{a+a^{-1}}\biggr)^{l+3/2} \biggl(\frac{1-a^2}{1+a^2}\biggr)^{n+n'} \sum_{k=0}^{n_<} \frac{(-1)^k}{k!}
    \notag\\&\phantom{{}={}}\times
  \frac{\sqrt{n!n'!}}{(n-k)!(n'-k)!} \frac{\sqrt{\Gamma(n+l+3/2)\Gamma(n'+l+3/2)}}{\Gamma(k+l+3/2)}
    \notag\\&\phantom{{}={}}\times
  \biggl(\frac{2a}{1-a^2}\biggr)^{2k},
\end{align}
where $n_<=\min(n,n')$.
The second equality shows that the overlap possesses a high degree of symmetry:
\begin{align}
  I_l(nb,n'b') &= (-1)^{n+n'} I_l(n'b,nb') \\
    &= (-1)^{n+n'} I_l(nb',n'b) \\
    &= I_l(n'b',nb)
\end{align}
The first relation arises because the only asymmetry between $n$ and $n'$ is the phase factor.
To get the second, we note that exchanging the oscillator lengths replaces $a$ by its reciprocal, which only appears in symmetric expressions or fractions that can be returned to their original form by expanding with a power of $a$.
The only change is a factor of $(-1)^{n+n'}$ from the third term.

\renewcommand{\doibase}[0]{https://doi.org/}%
\newcommand{\Eprint}[0]{\hfil\penalty100\hfilneg\href }

\end{document}